\documentclass[twocolumn]{aa} 
\usepackage{graphicx}
\usepackage{natbib}
\bibpunct{(}{)}{;}{a}{}{,} 
\usepackage{txfonts}
\begin{document}
   \title{Metallicity of high stellar mass galaxies with signs of merger events}

   \subtitle{}

   \author{M. Sol Alonso\inst{1,2}
          \and
          Leo Michel-Dansac\inst{1,3,4}
          \and
          Diego G. Lambas\inst{1,3}
          }

   \institute{Consejo Nacional de Investigaciones Cient\'{\i}ficas y
     T\'ecnicas, Argentina\\ \email{salonso@icate-conicet.gob.ar,
       leo@mail.oac.uncor.edu, dgl@mail.oac.uncor.edu} 
     \and Instituto de Ciencias Astron\'omicas, de la Tierra y del
     Espacio, San Juan, Argentina
     \and IATE, CONICET, OAC, Universidad Nacional de C\'ordoba,
     Laprida 854, X5000BGR, C\'ordoba, Argentina
     \and Centre de Recherche Astrophysique de Lyon, Universit\'e de Lyon,
     Universit\'e Lyon 1, Observatoire de Lyon, Ecole Normale Sup\'erieure de
     Lyon, CNRS, UMR 5574, 9 avenue Charles Andr\'e, Saint-Genis Laval, 69230,
     France }

   \date{Received ; accepted }


  \abstract
   {}
{We focus on an analysis of galaxies of high stellar
  mass and low metallicity. Evidence of recent merger events
  in their optical images allow us to classify galaxies
  into either disturbed or undisturbed, and to study galaxy properties such
  as morphology, colours, stellar populations, and global environment 
  for the different metallicity ranges and disturbance classes.}
{We cross-correlated the Millenium Galaxy Catalogue (MGC) and the Sloan
  Digital Sky Survey (SDSS) galaxy catalogue to
  provide a sample of MGC objects with high resolution imaging and
  both spectroscopic and photometric information available in the SDSS database. For
  each galaxy in our sample, we conducted a systematic
  morphological analysis by visual inspection of MGC images using
  their luminosity contours. The galaxies are classified as either 
  $disturbed$ or $undisturbed$ objects.  We divide the sample into
  three metallicity regions, within wich we compare the 
  properties of disturbed and undisturbed objects.}
{We find that the fraction of galaxies that are strongly disturbed, 
  indicative of being merger remnants, is higher when lower
  metallicity objects are considered.  The three bins analysed consist
  of approximatively 15\%, 20\%, and 50\% disturbed galaxies (for high, medium, 
  and low metallicity, respectively).  Moreover, the ratio of the disturbed to
  undisturbed relative distributions of the population age indicator,
  D$_n$(4000), in the low metallicity bin, indicates that the disturbed objects have
  substantially younger stellar populations than their undisturbed
  counterparts.  In addition, we find that an analysis of colour distributions
  provides similar results, showing that low metallicity galaxies with
  a disturbed morphology are bluer than those that are undisturbed. The
  bluer colours and younger populations of the low metallicity,
  morphologically disturbed objects suggest that they have experienced
  a recent merger with an associated enhanced star formation rate. }
{}

   \keywords{galaxies: formation - galaxies: evolution - galaxies:
abundances - galaxies: interactions.
}

   \maketitle
%

\section{Introduction}

In the most commonly accepted cosmological paradigm, galaxy
interactions and mergers play a crucial role in determining galaxy
properties and are considered as one of the main mechanisms by which
galaxies experience significant changes in morphology, stellar
population content, and star formation activity
\citep[e.g.,][]{BGK00,LTAC03,ALTC06}.  Thus, it is generally accepted
that the different merger histories of galaxies define their
evolution, origin, and present day properties.

It is also important to consider the chemical properties of galaxies.
These can provide fossil records of their history of formation
\citep{FBH02} since the metallicity content of a galaxy is expected to
depend strongly on its evolutionary state, namely, how and when was
the gas transformed into stars.  The relation between
interactions/mergers and chemical properties have been studied by
different authors \citep{DP00,MMMVBG02,Fetal04,KGB06,MLAT}. The
underlying idea in these studies has been that a close companion can
induce gas inflows which lower the metallicity in the central regions
of galaxies \citep{KGB06,MLAT}.

A stellar mass-gas metallicity relation has been well established
\citep[hereafter MZR,][]{Tetal04}. Although it is a clearly observed
trend of increasing gas-phase metallicity with stellar mass, there is
considerable scatter in the relation, which could be attributed to the
particular star formation history of galaxies. Accepted theories for
the origin of the MZR have as a central proposition that efficient
galactic outflows remove metals from galaxies with shallow potential
wells \citep{LA74,Dal07,FD08}. In this context, mergers and
interactions could play an important role in determining the shape and
scatter of the MZR.

\citet{KGB06} detected a shift in the luminosity - metallicity (LZ)
relation towards lower metallicities by $\approx$ 0.2 dex for galaxy
pairs of a given luminosity, compared to a control sample. The spectra
analysed in this work corresponded only to the central 10$\%$ of the
galaxy and the authors interpreted this result as a signature of
metal-poor gas being funneled into the center of the galaxies.
Following this approach, Ellison et al. (2008) studied the LZ relation
of 1716 galaxies with companions with radial velocity $\Delta$V $<$
500 km.s$^{-1}$ and projected separation rp $<$ 80 kpc h$^{-1}$, and
they found an offset to lower metallicities (by $\approx$ 0.1 dex) for
a given luminosity in pair galaxies. \citet{MLAT} studied the stellar
mass - gas metallicity relation of galaxies in close pairs
(morphologically classified according to the strength of the
interaction signatures) and in isolation taken from the Sloan Digital
Sky Survey Data Release 4 (SDSS-DR4).  The authors measured
differences in the metallicity of galaxies at a given stellar mass,
between pairs showing signs of strong interactions and galaxies in
isolation. Thus, the mass-metallicity relation differs between systems
with a companion and galaxies in isolation.  Interacting galaxy pairs
of high stellar mass are found to have systematically lower
metallicity values than the mean MZ relation, which implies that gas
inflows have been induced by the interaction.  In addition,
\citet{PPS} studied 42 outliers (massive, metal-poor galaxies) of the
MZR in the SDSS, finding that these objects are in general extremely
blue and star-forming, and show signs of morphological disturbances.

On larger scales, the MZR appears to be only a weak function of galaxy
environment \citep{mouhcine+07,Cooper+08}.  The gas-phase oxygen
abundance of galaxies (at a given stellar mass) in the richest
environments is only $\sim 0.05$ dex higher than those in the poorest
environments, and these trends might account for about $15\%$ of the
scatter in the MZR.  \citet{Ellison+09} explored the impact of cluster
membership and local-density on the MZR, confirming that galaxies in
clusters are slightly more metal-rich that in the field but that this
effect is driven by local overdensity and not simply cluster
membership.

Given that the majority of galaxies lie close to the best-fit
mass-metallicity relation, we may be able to learn about the gas-phase
metallicity evolution of galaxies by studying the properties of
galaxies that do not follow this relation.  Therefore, joint
statistical analysis of mergers/interactions and chemical properties
could help us to shed light on the efficiency and extent of the
processes mentioned above.  To this aim, we consider star-forming,
high-stellar-mass galaxies that have in general a well-defined
morphology in contrast to smaller systems. It is therefore much easier
to identify a recent merger event in these objects.  We study high
resolution $B-$band images of high mass ($10< \log M_*/M_{\sun} <
10.8$) galaxies extracted from the Millennium Galaxy Catalogue
\citep[hereafter MGC,][]{MGC1,MGC5}.  We classify the galaxies as
either disturbed or undisturbed and we explore statistically the
effects of recent merger events on oxygen abundance, colour, and
stellar age indicators derived from SDSS.


\section{Data}

\begin{figure}
  \includegraphics[width=84mm]{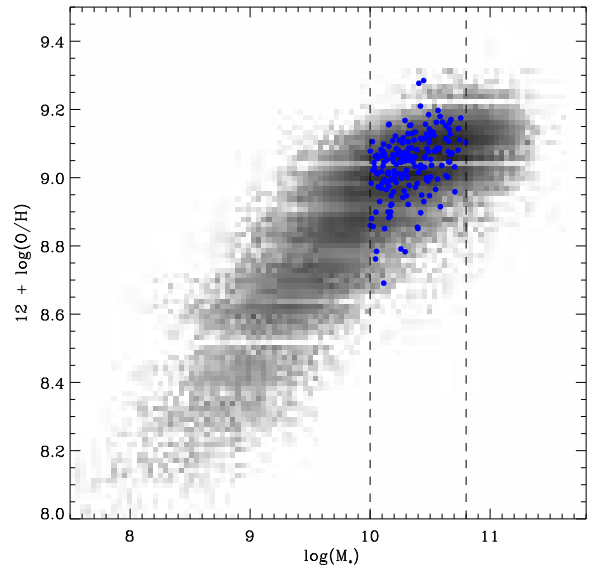}
  \caption{Stellar mass - gas metallicity relation of a selection of
    the SDSS galaxies (see text). The vertical dotted lines delimit
    the mass range that we study in detail in this paper.  The blue
    points correspond to our sample of MGC-SDSS galaxies.  }
  \label{mzrmorfo}
\end{figure}

Our data consist of galaxies from the Millennium Galaxy Catalogue
\citep{MGC1, MGC5} and the Sloan Digital Sky Server Data Release 4
\citep{sdssdr4}.

The Millennium Galaxy Catalogue is a wide medium-deep $B-$band imaging
survey along the celestial equator that was obtained using the Wide
Field Camera on the 2.5-m Isaac Newton Telescope on La Palma.  The
survey covers 37.5 square degrees with $B-$band magnitudes in the
range 13 $< B <$ 24.  The survey region coincides with both the 2dF
Galaxy Redshift Survey northern strip \citep[2dFGRS, ][]{2df} and the
Sloan Digital Sky Survey region \citep{sdss}.

The SDSS is a photometric and spectroscopic survey that will cover
approximately one-quarter of the celestial sphere and collect spectra
for more than one million objects.  The imaging portion of the fourth
release SDSS comprises 6670 square degrees of sky imaged in five
wavebands ($u$, $g$, $r$, $i$, and $z$) containing photometric
parameters of 180 million objects.  The spectra are obtained from 3
arcsec diameter fibers projected on the sky and the coverage is
3800-9200 ${\AA}$.

The SDSS value-added catalog provides several spectroscopic parameters
of physical relevance.  However, SDSS imaging is relatively shallow
and often taken in poor seeing conditions.  Therefore, because we wish
to classify galaxies according to their degree of disturbed morphology
we use a subset of SDSS galaxies in common with MGC, which provides
deeper and higher resolution imaging data.  The quality of the images
in MGC is a major asset for a detailed study of galaxy morphology. The
median seeing of the MGC survey, 1.3 arcseconds \citep{MGC1}, combined
with its deep surface brightness limit of $\mu_B = 26$ mag
arcsec$^{-2}$, provides much higher quality and more clearly resolved
images than SDSS images of surface brightness limit 20 mag
arcsec$^{-2}$, which were obtained with a median seeing of 1.5
arcseconds.

We cross-correlated the MGC and SDSS DR4 galaxy catalogs to obtain a
sample of MGC objects with oxygen abundances
$12+\log(\ion{O}/\ion{H})$ \citep{Tetal04}, stellar mass $M_*$, and
$4000$-\AA\ break strength D$_n$(4000) \citep{Ketal03}, as well as
basic information in SDSS data (e.g., redshifts, concentration
parameter).

We characterized the local environment of galaxies by defining a
projected density parameter, $\Sigma_5$. This parameter is calculated
by using the projected distance to the $5^{th}$ nearest neighbour,
$\Sigma_5= 5/(\pi d_{5}^{2})$.  Neighbours were chosen to have
luminosities brighter than $M_{r}<-19.5$ and radial velocity
differences lower than $1000$ km.s$^{-1}$ \citep{BAL04}.
\citet{Ketal04} estimated the local density by counting galaxies
within cylinders of 2 Mpc in projected radius and $\pm$ 500
km.s$^{-1}$ in depth, in a complete sample of galaxies from SDSS.
\citet{ALTC06} found a good correlation signal between both density
estimators indicating that either of these two definitions are
adequate for characterizing the environment of galaxies.  Studies of
the relationship between gas-phase oxygen abundance and environment
have used similar parameters, for instance the $\Sigma_3$ projected
density parameter \citep{Cooper+08}, or a density estimator based on
the average of the projected distances to the fourth and fifth nearest
neighbour within $1000$ km.s$^{-1}$ \citep{mouhcine+07}.

We imposed an upper redshift cutoff at $z<0.1$ to ensure both high
completeness for both the SDSS and MGC data, and sufficiently high
angular resolution. Additionally, we considered a minimum redshift
$z>0.04$ to avoid a too low spatial coverage by the SDSS fibers.
Thus, for the range adopted ($0.04<z<0.1$) the 3 arcsec SDSS fiber
probes approximatively the central 2.5 kpc to 6 kpc of the galaxies.

We restricted the present analysis to high stellar mass objects in the
range $10< \log M_*/M_{\sun} < 10.8$.  The high stellar mass range
adopted is consistent with rather extended galaxies so that
metallicity measures do not include the external parts of the
galaxies.  Moreover, the rather narrow stellar mass range adopted
ensures that our subsample of galaxies is homogeneous and adequate for
the detection of substructures.

To check for any potential AGN contamination in our sample, we
analysed the BPT diagnostic diagrams $[\ion{O}{iii}]/\ion{H}{\beta}$
versus (vs.) $[\ion{N}{ii}]/\ion{H}\alpha$, and both
$[\ion{O}{iii}]/\ion{H}\beta$ vs. $[\ion{O}{i}]/\ion{H}\alpha$ and
$[\ion{O}{iii}]/\ion{H}\beta$ vs. $[\ion{S}{ii}]/\ion{H}\alpha$
proposed by \citet{BPT81}, in all cases using the classification
criteria given in \citet{K06}.  We found that a small percentage of
the objects have line ratios indicative of their spectra being
possibly associated with composite AGN/$\ion{H}{ii}$ regions.  To
avoid this possibility, we removed these ambiguous cases from the
analysis.

A possible concern with SDSS spectroscopy is the signal-to-noise ratio
of the relevant lines associated with gas-phase oxygen abundance
determinations.  We checked that in our sample all galaxies have a
signal-to-noise ratio (S/N) of at least 5 in the lines H$\alpha$,
H$\beta$, and $[\ion{N}{ii}]\lambda6584$, and at least 3 in the
$[\ion{O}{iii}]\lambda5007$ line following the selection criteria of
\citet{Tetal04}.  Thus, despite metallicity determination being a
complex task with a rather large associated uncertainty, our analysis
is not likely to be affected by systematic biases (see also discussion
in Sect.~\ref{part32}).

\begin{figure}
  \includegraphics[width=84mm]{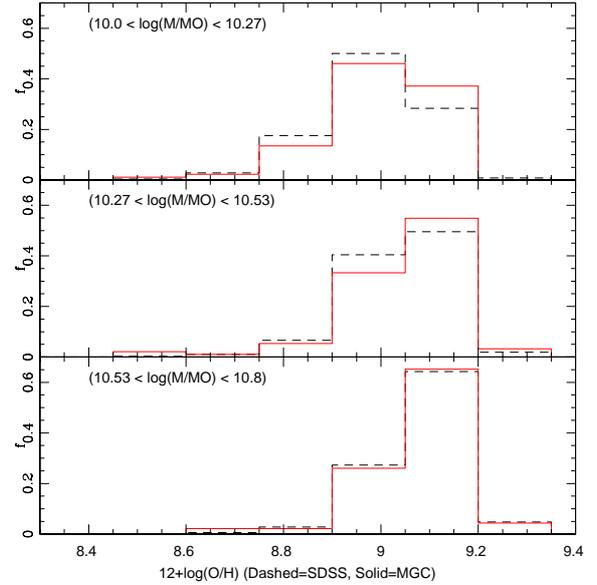}
  \caption{Oxygen abundance distributions: dashed lines correspond to
    the SDSS Tremonti et al. data, and solid lines to our sample of
    MGC-SDSS galaxies in three stellar mass bins.}
  \label{F1new}
\end{figure}

With the above restrictions, our final sample comprises 191 high
stellar mass galaxies.  In Fig.~\ref{mzrmorfo}, we plot the MZR for
the total sample of SDSS galaxies in the redshift range $0.04<z<0.1$
satisfying the selection criteria adopted (e.g., AGN removal, S/N),
and our sample of high stellar mass MGC-SDSS galaxies.
Figure~\ref{F1new} shows the $12+\log(\ion{O}/\ion{H})$ distribution
in both catalogs in 3 mass bins within the mass range selected
previously.  It can be appreciated that the MZR of our MGC-SDSS sample
is truly representative of a MZR determined for the whole SDSS in the
selected mass range.

\section{Analysis}

\subsection{Visual classification of galaxies into disturbed and
undisturbed objects}

\begin{figure*}
  \centering
  \includegraphics[width=180mm]{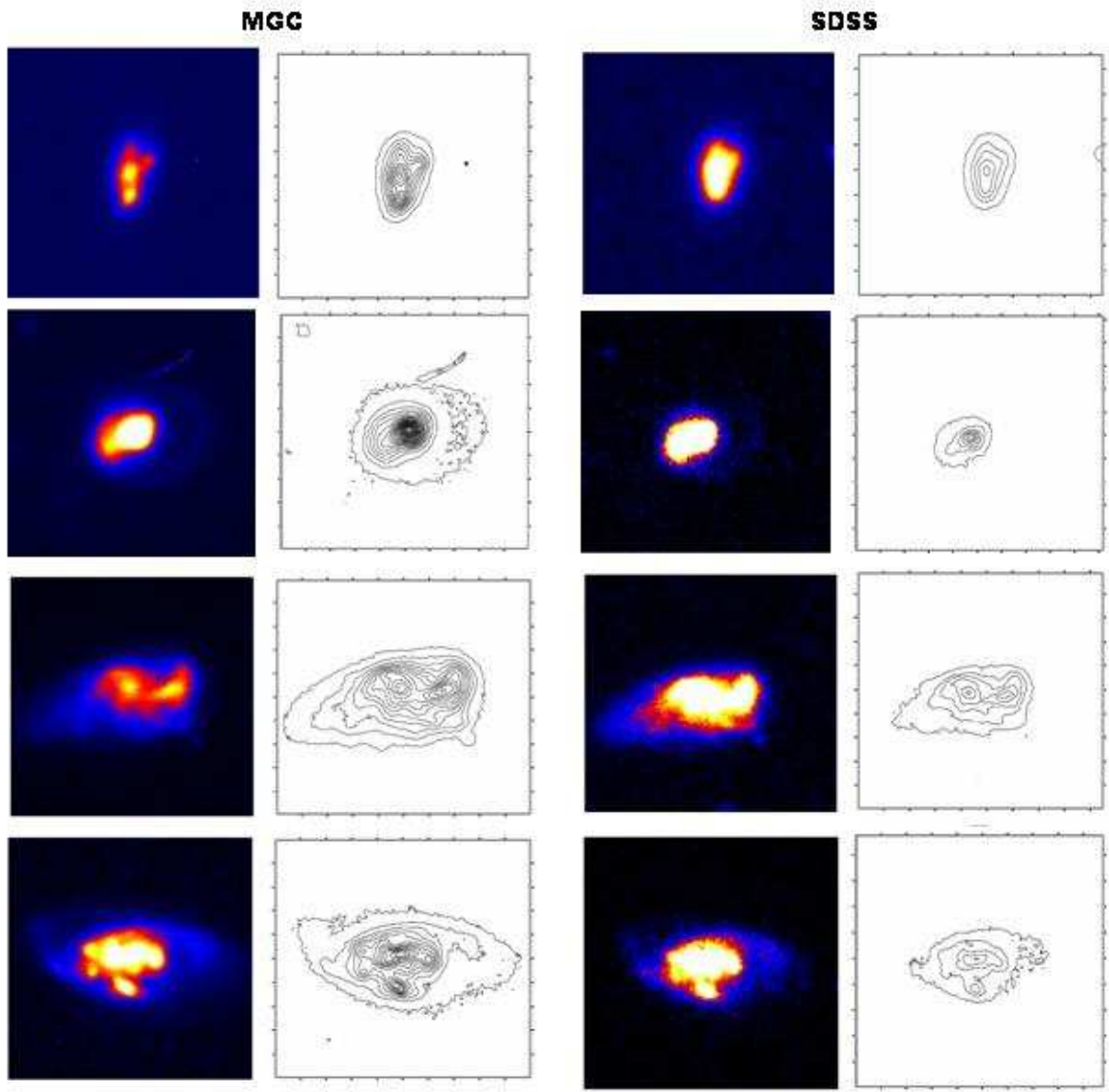}
  \caption{Images and surface brightness contours of typical massive
    disturbed galaxies in both MGC (left columns) and SDSS (right
    columns) catalogs.  The image sizes are 33 x 33 arcsec$^2$.  }
  \label{contours}
\end{figure*}

For each galaxy in our sample we conducted a systematic morphological
analysis by visual inspecting MGC images and studying the luminosity
contours obtained with IRAF routines applied to the reduced images
provided in the catalogue.  Galaxies are classified as:

$a$) $Disturbed$: these objects show externally-triggered distortions
indicative of a recent strong tidal interaction or merger. Some of
these objects exhibit arcs, shells, ripples, tidal debris, warps,
rings, extremely asymmetric light distributions, tidal tails, multiple
nuclei, and strong substructure inside a common body. We note that
these galaxies are not two interacting systems in a ongoing merger,
but a single galaxy with signs of being a merger remnant.

$b$) $Undisturbed$: objects with smooth luminosity contours.

To assess the advantage of using MGC images for the present analysis,
we compare in Fig.~\ref{contours} $B-$band MGC and $g-$band SDSS
images for a sample of disturbed galaxies where it is clearly apparent
that there is higher reliability when assigning a disturbed morphology
in the higher quality MGC data. It is clear that several cases would
not be classified as disturbed in SDSS, so that the improvement
provided by using MGC data is evident.  We also used SDSS data to
analyse the close local environment of each galaxy, since MGC images
have a too small field of view.  Since we are interested in analysing
the effects of past merger events, we excluded a few potentially
interacting pairs in the sample.

We considered the uncertainty in the visual classification of galaxy
morphology by independent determination by different observers. By
cross-checking the results, we estimate a 95\% coincidence indicating
the reliability of our classification scheme.

Visual classification is by nature a subjective task. To assess the
reliability of our classification, we derived the asymmetry $A$
parameter \citep{conselice+00} for the MGC images.  Different authors
\citep[e.g.,][]{dp07,jog+09} have shown that pairs and galaxy mergers
exhibit larger asymmetries.  In our sample, for each galaxy, we use
SExtractor \citep{sextractor} to calculate the $A$ parameter, finding
that galaxies classified as $disturbed$ by visual inspection have a
larger asymmetry parameter ($A \ga $ 0.30), and $undisturbed$ objects
have lower asymmetry values ($A \la$ 0.30). More precisely, we find
that $72 \%$ of disturbed galaxies exhibit $A \ga $ 0.3, whereas all
the galaxies classified as undisturbed have $ A \la $ 0.3.

\begin{figure}
  \includegraphics[width=84mm]{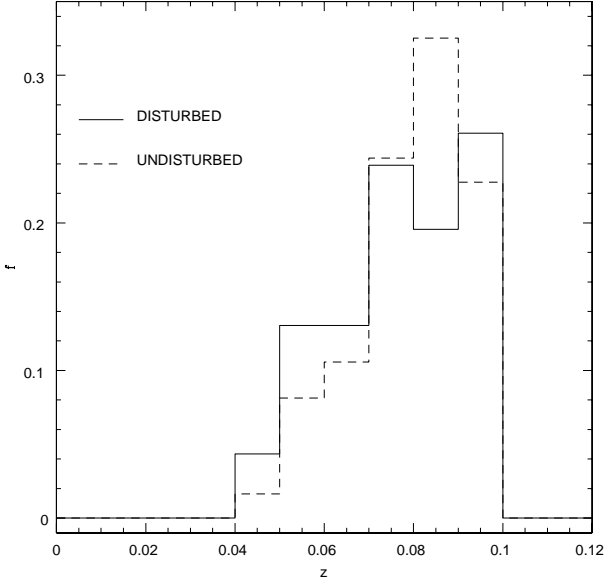}
  \caption{Redshift distribution of disturbed (solid line) and
    undisturbed galaxies (dashed line).}
  \label{histz}
\end{figure}

For the stellar mass range explored, we found 46 strongly
morphologically disturbed galaxies, 123 undisturbed galaxies, and 22
galaxies with a close companion.  The redshift distributions of
disturbed and undisturbed galaxies is shown in Fig.~\ref{histz}. The
great similarity between these distributions is an indication that our
classification is not affected by angular resolution.

\subsection{Oxygen abundances, colours, and star formation histories of disturbed and
  undisturbed galaxies}
\label{part32}

\begin{figure}
  \includegraphics[width=84mm]{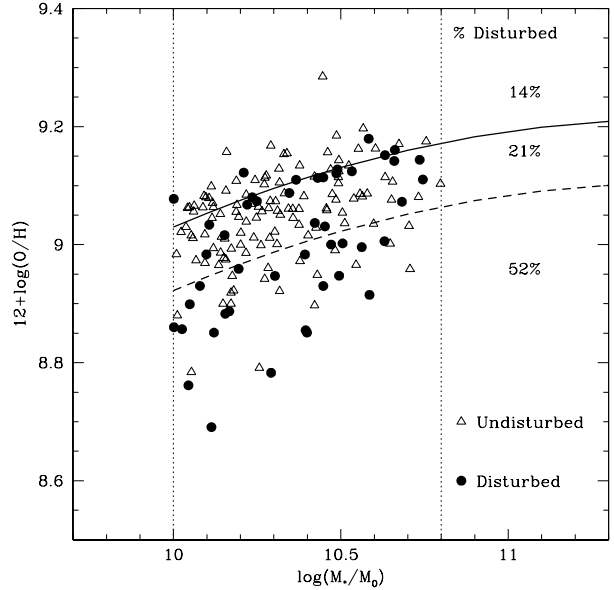}
  \caption{Distribution of $12+\log(O/H)$ and $\log(M_*/M_{\sun})$
    values for galaxies with stellar masses in the range
    $\log(M_*/M_{\sun})$ $\approx$ 10.0 to 10.8.  The circles
    correspond to disturbed galaxies and triangles to undisturbed
    galaxies.  The solid and dashed lines indicate the limits of the
    high, medium, and low metallicity ranges.  The solid line
    corresponds to the Tremonti et al. (2004) fit to the MZ relation
    shifted by +0.078. The dashed line corresponds to this fit shifted
    by -0.03 dex.}
  \label{mzrclas}
\end{figure}

\begin{figure}
  \includegraphics[width=84mm]{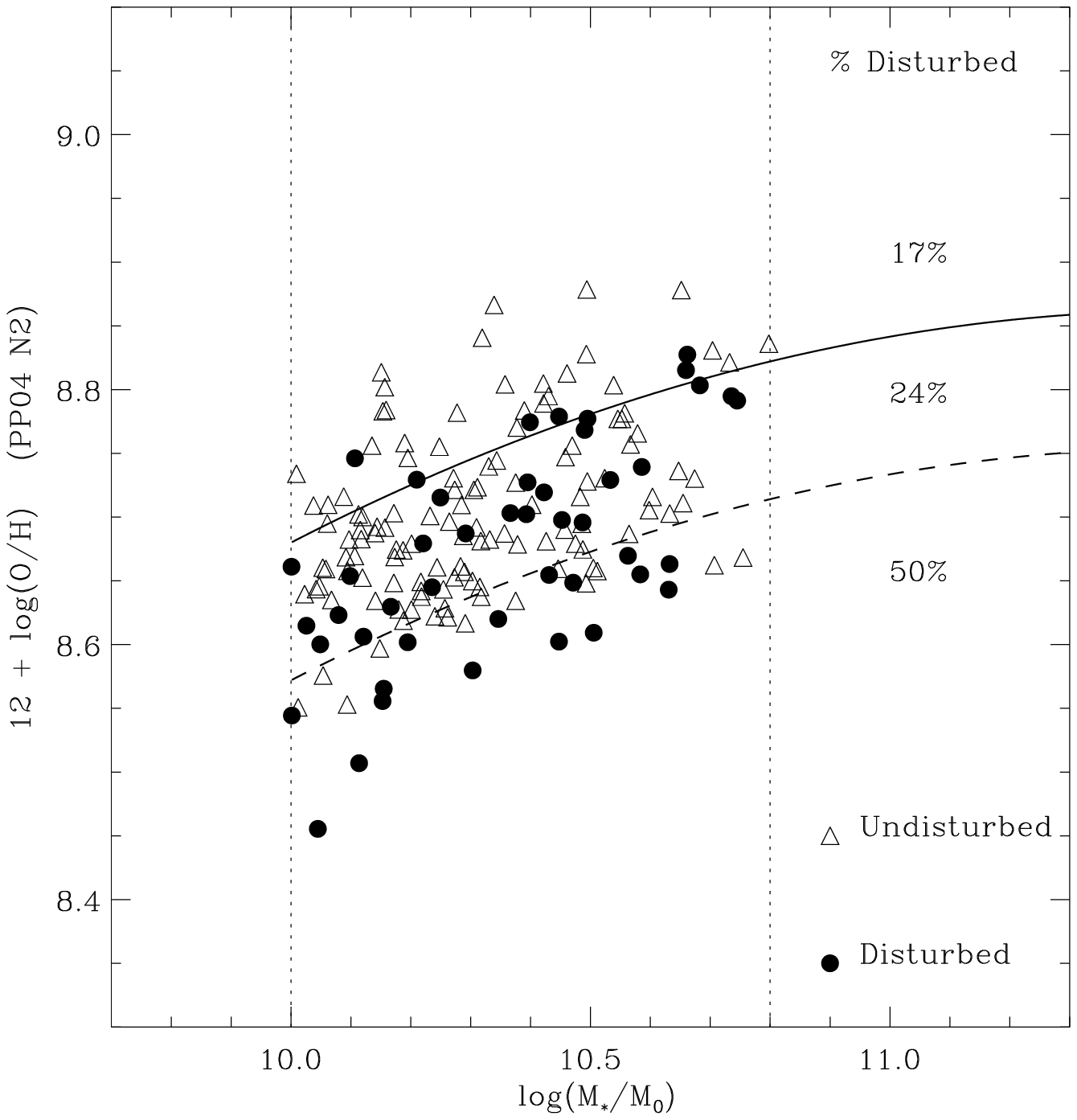}
  \caption{Distribution of $12+\log(O/H)$ calculated through the
    Pettini \& Pagel (2004) method, and $\log(M_*/M_{\sun})$ values
    for galaxies with stellar masses in the range $\log(M_*/M_{\sun})$
    $\approx$ 10.0 to 10.8. Symbols are the same as in
    Fig.~\ref{mzrclas}. The solid and dashed lines indicate the limits
    of the high, medium, and low metallicity ranges. With these
    abundances, the solid line corresponds to the Tremonti et
    al. (2004) fit to the MZ relation shifted by $-0.272$. The dashed
    line corresponds to this fit shifted by $-0.38$ dex.}
  \label{mzrclas2}
\end{figure}

We analysed the oxygen abundance and stellar mass in our sample of
high mass MGC-SDSS galaxies. We show the MZR of these galaxies in
Fig.~\ref{mzrclas}.  We considered two threshold curves displaced with
respect to the Tremonti et al. fit, which separate the sample into
three regions with low, medium, and high metallicity defined by:

high: $12+\log(\ion{O}/\ion{H}) > -1.492+1.847(\log M_*)-0.08026(\log
M_*)^2$+0.078;

low:  $12+\log(\ion{O}/\ion{H}) < -1.492+1.847(\log M_*)-0.08026(\log
M_*)^2$-0.03;

and a medium metallicity range between these two. Adopted thresholds
divide the sample into approximatively 1/4, 2/4, and 1/4 the total
sample of galaxies (representing the high, medium, and low metallicity
ranges respectively).

\begin{table*}
  \caption{Number of galaxies classified as disturbed galaxies,
    undisturbed galaxies, and galaxies with a close companion pairs,
    in our three ranges of metallicity, and percentages (and
    respectives sampling errors) of disturbed galaxies with respect to
    the total number of disturbed and undisturbed galaxies} 
  \centering
  \begin{tabular}{lccccc}
    \hline\hline
    &  disturbed  &  undisturbed  &  with close comp. & total & $\%$ disturbed \\
    \hline
    total    &   46  & 123   & 22 & 191 & $27.2 \pm 3.4$ \\ 
    \hline
    high O/H   & 6  & 35  &  6 &  47    & $14.6 \pm 5.8$ \\
    medium O/H & 18 &  68 & 13 &  99    & $20.9 \pm 4.4$ \\
    low O/H    & 22 &  20 &  3 &  45    & $52.3 \pm 7.7$ \\
    \hline
  \end{tabular}
  \label{numbers}
\end{table*}

As shown in Fig.~\ref{mzrclas} and Table~\ref{numbers}, the relative
fractions of disturbed galaxies are statistically significantly
different in the three metallicity bins explored. This remarkable
tendency for the relative number of disturbed galaxies to increase
with decreasing metallicity is also evident from visual inspection of
Fig.~\ref{mzrclas}.  To quantify these results, we also calculated the
percentages of disturbed galaxies for the three different metallicity
bins defined previously. At medium and high metallicities, we find
that the percentages of disturbed objects are not very different (14\%
and 21\% respectively, cf Table~\ref{numbers}).  However, at low
metallicities, the disturbed galaxy fraction increases remarkably
(52\%).

To check for possible errors in the metallicities of galaxies in our
sample, we recalculated $12+\log(\ion{O}/\ion{H})$ using the
\citet{PP04} method, where the oxygen abundance is given by
$12+\log(\ion{O}/\ion{H})=9.37+2.03\times\mathrm{N}2+1.26\times\mathrm{N}2^2
+0.32\times\mathrm{N}2^3$ and $\mathrm{N}2 = \log
([\ion{N}{ii}]\lambda6584 / \ion{H}\alpha)$.  This method is
particularly suited to our sample since all galaxies have a S/N at
least 8 in the emission lines used to compute N2. Moreover, the N2
range of the galaxies studied ($-0.7 < \mathrm{N}2 < -0.29 $) is
within the domain of validity for this equation.  We considered the
three metallicity regions defined previously but shifted by $-0.35$
dex since this method tends to infer lower values of metallicities
than those of \citet{Tetal04} \citep[see for instance][]{KE08}. With
these new estimates, we obtain similar numbers of galaxies in each
metallicity bins, as shown in Fig.~\ref{mzrclas2}.  We also used a
subsample of 82 galaxies with data of S/N of at least 8 in the strong
emission lines and oxygen abundances measured by \citet{Tetal04},
finding very similar percentages (63, 22, and 17\%) of disturbed
galaxies in the low, medium, and high metallicity bins. Then, this
trend of increasing the disturbed galaxy fraction at lower metallicity
appears robust and is not likely to be caused by errors in the
determination of oxygen abundances for these galaxies.

\begin{figure}
  \includegraphics[width=84mm]{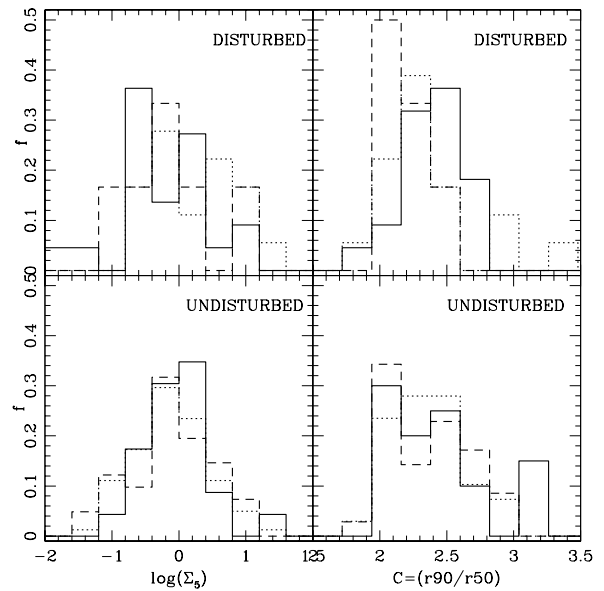}
  \caption{Distribution of the environmental local galaxy density
    parameter $\Sigma_5$ and image concentration parameter $C$ for
    disturbed and undisturbed objects in the three metallicity ranges:
    low (solid), medium (dotted), and high (dashed line).  }
  \label{sigcon}
\end{figure}

Studies of galaxy properties and their dependence on environment are
important for understanding the role of mechanisms driving the
evolution of galaxies.  We therefore explored the possible influence
of the environment of each galaxy in our sample on the results.  To
this aim, we analysed the local density parameter $\Sigma_5$, defined
previously, for all galaxies in our sample.  The results of this
analysis are shown in Fig.~\ref{sigcon}, where it can be appreciated
that disturbed and undisturbed galaxies populate similar environments
indicating that the effects are not caused by the particular location
of these systems.  Thus, we conclude that the strong trend of
increasing percentages of disturbed galaxies as a function of
decreasing metallicity is not likely to be biased by environmental
effects.

For completeness in our analysis, we also studied the light
concentration parameter (i.e., the ratio of Petrosian 90\% to 50\%
$r-$band light radii, $C=r_{90}/r_{50}$) of the galaxies in the sample
to search for possible differences in our sample between disturbed and
undisturbed objects.  The concentration index value of $C=2.5$ was
adopted to segregate concentrated, bulge-like ($C >2.5$) galaxies from
more extended, disc-like ($C<2.5$) systems.  In Fig.~\ref{sigcon}, the
similarity between the $C$ parameters across the samples is clearly
evident, showing that the observed disturbed to undisturbed ratio
dependence on metallicity is probably not associated with systematic
differences in the bulge-to-disc ratios of the galaxies.  To quantify
the significance of the differences between the distributions in
Fig.~\ref{sigcon}, we calculated the occurrence of $C<2.5$ in galaxies
with low metallicities in 20 randomized samples.  In each sample, the
$C$ parameter was reassigned randomly, in disturbed and undisturbed
objects of low metallicity.  Since we found that in 6 out of 20
randomized samples the signal is as strong as in the data, we conclude
that no statistically significant tendency associated with disturbed
and undisturbed galaxies can be identified for the concentration
parameter.

\begin{figure}
  \includegraphics[width=84mm]{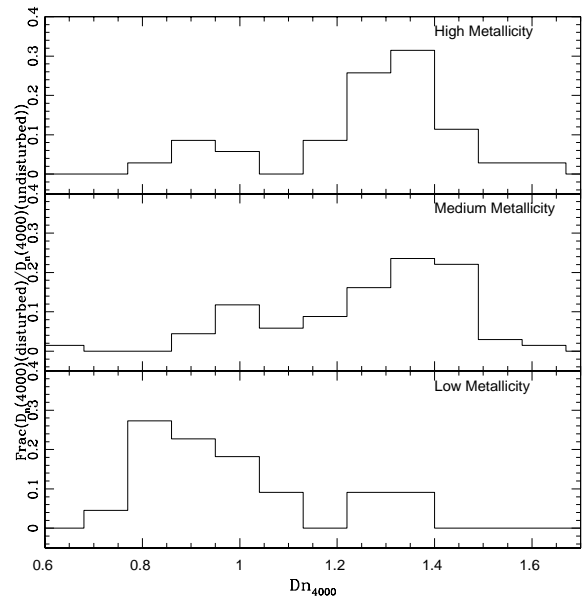}
  \caption{Fraction of stellar age indicator (D$_n$(4000)) for
    disturbed normalized to undisturbed galaxies
    (D$_n$(4000)($disturbed$)/D$_n$(4000)($undisturbed$)) in the three
    metallicity ranges.}
  \label{Dn}
\end{figure}

That the relative fraction of disturbed to undisturbed galaxies
increases for low metallicity objects provides important evidence that
a merger process induces low metallicity gas inflows. This phenomenon
should also be reflected in the mean stellar age population of
disturbed objects as a function of metallicity.  The D$_n$(4000)
spectral index defined by \citet{bruzual83} is a suitable indicator of
the current luminosity-weighted mean stellar age since young stellar
populations produce almost no metal absorption just shortward of
$4000\AA$. Therefore, galaxies dominated by recent episodes of star
formation exhibit a small index, while the spectra of galaxies
dominated by older populations exhibit strong metal lines in
absorption and thus larger D$_n$(4000) values. We analysed the
D$_n$(4000) population age indicator in the 3 gas-phase oxygen
abundance ranges.  The results are shown in Fig.~\ref{Dn}, where we
plot the fraction of D$_n$(4000) for disturbed objects normalized to
that for undisturbed galaxies
(fraction(D$_n$(4000)($disturbed$)/D$_n$(4000)($undisturbed$))).  We
find that at both medium and high metallicity, both disturbed and
undisturbed objects have a similar D$_n$(4000), with a slight tendency
for disturbed objects to have older stellar populations.  It can also
be appreciated that low O/H values objects, where the disturbed
population is important, have substantially younger stellar
populations than their undisturbed counterparts. Thus, our
morphological analysis indicates that in high stellar mass galaxies,
low metallicity values are associated with star formation episodes
triggered by recent merger events.  By adopting the same procedure
used previously to quantify the differences between the distributions
of the $C$ parameter, we test the differences between the D$_n$(4000)
distributions shown in Fig.~\ref{Dn}. Here the results show high
confidence in the significance of the differences since none of the 20
randomized realizations exhibit a signal as strong as in the data. We
conclude that at low metallicities, disturbed galaxies have a younger
stellar population at least at the $95\%$ confidence level.

\begin{figure}
  \includegraphics[width=84mm]{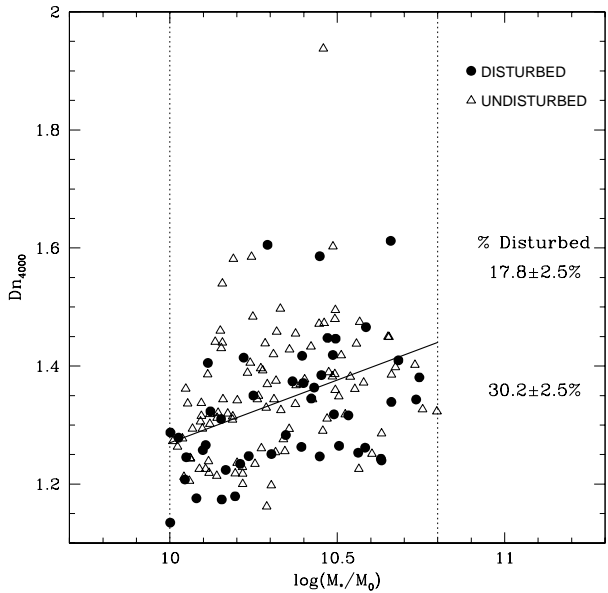}
  \caption{Distribution of D$_n$(4000) and $\log(M_*/M_{\sun})$ values
    for galaxies with stellar masses in the range $\log(M_*/M_{\sun})$
    $\approx$ 10.0 to 10.8.  The circles correspond to disturbed
    galaxies and triangles to undisturbed galaxies.  }
  \label{dnmstn}
\end{figure}

We also considered the fraction of disturbed galaxies above the median
D$_n$(4000) for a given value of $M_*$.  By doing so, we aim to relate
merging events to the relative fraction of young stellar populations.
In Fig.~\ref{dnmstn}, we show the observed D$_n$(4000) vs $M_*$ for
disturbed and undisturbed galaxies. As can be appreciated in this
figure (see quoted numbers), there is a larger fraction of disturbed
galaxies with stellar population ages above the median defined by
D$_{n}(4000) = 0.2125 \log(M_*/M_{\sun}) - 0.85$ (shown as a solid
line).  Thus, our classification of disturbed type selects
preferentially galaxies with the highest fraction of young stellar
populations at a given stellar mass content, as expected in a
merger-induced star formation scenario.  We addressed the
uncertainties in our estimated fractions of disturbed and undisturbed
galaxies by moving the median line threshold of 0.3 dex. We find that
the relative fractions do not change significantly indicating that the
results are not sensitive to the particular choice adopted.

\begin{figure}
  \includegraphics[width=84mm]{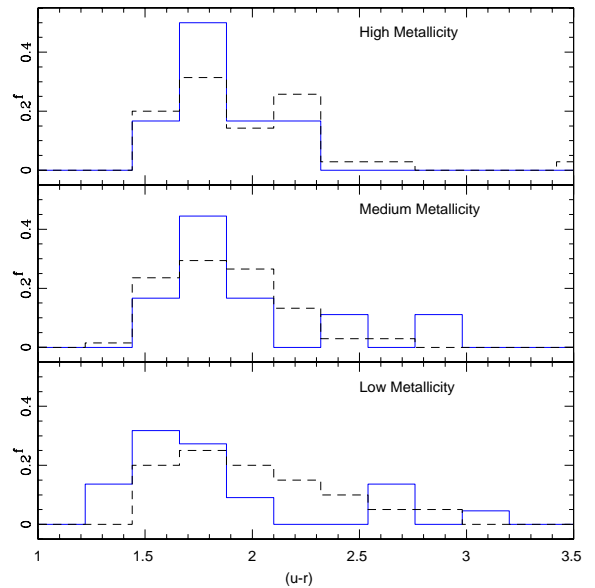}
  \caption{Colour index ($u-r$) for disturbed (solid) and undisturbed
    galaxies (dashed) in the three adopted metallicity ranges. }
  \label{col}
\end{figure}

Observational analyses of the $u-r$ colour distributions
\citep[e.g.,][]{ALTC06, dp05} have reported an excess of blue colours
in close pair galaxies with signs of interactions, relative to those
measured for galaxies without close companions.  This finding is
associated with both a larger fraction of actively star-forming
galaxies and younger stellar populations in disturbed close pair
systems.  In Fig.~\ref{col}, we compare the $u-r$ colour distributions
of disturbed galaxies with those of their undisturbed counterparts.
The analysis was performed for the three different metallicity ranges
previously considered (see Table~\ref{numbers}).  We find that in both
the high and medium metallicity ranges (upper and medium panels,
respectively), the colour distributions of disturbed objects are more
similar to those of undisturbed galaxies than in the lower metallicity
range.  Interestingly, at lower metallicity (lower panel), the colour
distributions of disturbed and undisturbed galaxies exhibit
significant differences, morphologically disturbed objects having a
larger fraction of blue colours.  We also observe a significant
difference in the blue tail, where there is a clear excess of blue
disturbed galaxies with respect to undisturbed objects.  The blue
excess can be interpreted in terms of substructures associated with
externally-triggered distortions, indicative of recent interactions or
mergers, and both enhanced star formation and young stellar
populations as already shown in Fig.~\ref{Dn}.

\section{Discussion and Conclusions}

Observed metallicity gradients in galaxies
\citep[e.g.,][]{zaritsky+94,ferguson+98,dutil99,considere+00} show
evidence of less enriched gas in the external regions. Thus, a large
low-metallicity gas reservoir is expected in most massive star-forming
galaxies. On the other hand, numerical simulations indicate that
galaxy interactions and mergers may generate a significant flow of
external gas onto the central region and trigger star formation. Thus,
interactions may be efficient in lowering the central metal content of
galaxies by means of an inflow of this external, less enriched gas.

Similar effects have been reported by \citet{MLAT} and \citet{EPSM08},
where interacting pairs with strong morphological disturbance features
have lower metallicities. This is also reinforced by the high mass
outliers in the MZ relation exhibiting distorted morphologies
\citep{PPS}.

The above considerations have led us to analyse the presence of
possible relics of such process in non-interacting, high-stellar-mass
galaxies. To this aim, we have performed a morphological analysis of
high quality images from the MGC of high-stellar-mass SDSS
galaxies. Objects in this mass range have a more clearly defined
morphology and so are suitable for exploring morphological
disturbances associated with merger events. On the other hand, the
oxygen abundance is a clear indicator of the gas phase metallicity and
so is sensitive to the presence of recently formed stars.

By construction, our subsamples of high, medium, and low metallicity
objects have similar stellar-mass distributions. Moreover, we have
confirmed that the subsamples are free of strong biases widely
different local galaxy density environments, or widely different
concentration parameters so that we expect them to have similar
bulge-to-disc ratios.  Thus, the subsamples analysed comprise galaxies
of similar morphology and environment.

Our main result is that we detect a steadily increasing fraction of
morphologically disturbed galaxies, characteristic of being merger
remnants, with decreasing O/H. The three bins analysed consist
approximatively of 15\%, 20\%, and 50\% disturbed galaxies (high,
medium, and low metallicity, respectively)

Moreover, the ratio of disturbed to undisturbed distributions of the
D$_n$(4000) parameter in both the medium and high metallicity bins are
similar, the D$_n$(4000) values corresponding to a relatively old
stellar population in disturbed objects. This contrasts with the
results for the low metallicity bin where disturbed objects have a
substantially younger stellar populations than their undisturbed
counterparts.  In addition, the colours distributions exhibit similar
trends, showing that low metallicity galaxies with a disturbed
morphology are bluer than those that are undisturbed.

The bluer colours and lower D$_n$(4000) values in low metallicity,
morphologically disturbed objects, suggest externally-triggered
distortions driven by a recent interaction or merger, with an
associated enhanced star-formation rate and a predominance of a young
stellar population.

\citet{RVB08} showed that ultra-luminous infrared galaxies (ULIRGs)
have systematically lower O/H values than star-forming galaxies of
similar stellar mass content, consistent with a merger-driven inflow
of low metallicity gas. Our results show, in addition, that this
mechanism is not only present in ULIRGs but also in high stellar mass
galaxies with morphological evidence of a merger event.

The most natural explanation of our results is that merger-induced low
metallicity gas inflows have occurred from the external regions of
high stellar mass galaxies. Our results also suggest that the
timescale of the chemical evolution (the combined effects of the
enrichment of the interstellar medium by SNII and these inflows) is at
least as long as the timescale of the tidally induced morphological
disturbances. Thus, galaxies with a disturbed appearence may contain
young stellar populations and yet low gas-phase abundances in their
central parts. Our results suggest that, at least in part, the scatter
in the MZR at high stellar mass is caused by interactions and mergers.

\section*{Acknowledgments}

We thank the referee for helpful comments and suggestions.  This work
was partially supported by the Consejo Nacional de Investigaciones
Cient\'{\i}ficas y T\'ecnicas, the Agencia de Promoci\'on de Ciencia y
Tecnolog\'{\i}a, the Secretar\'{\i}a de Ciencia y T\'ecnica de la
Universidad Nacional de C\'ordoba, and the MinCyT - ECOS-Sud program
\#A07U01.

The Millennium Galaxy Catalogue consists of imaging data from the
Isaac Newton Telescope and spectroscopic data from the Anglo
Australian Telescope, the ANU 2.3m, the ESO New Technology Telescope,
the Telescopio Nazionale Galileo and the Gemini North Telescope. The
survey has been supported through grants from the Particle Physics and
Astronomy Research Council (UK) and the Australian Research Council
(AUS). The data and data products are publicly available from
http://www.eso.org/~jliske/mgc/ or on request from J. Liske or
S.P. Driver.

Funding for the SDSS and SDSS-II has been provided by the Alfred
P. Sloan Foundation, the Participating Institutions, the National
Science Foundation, the U.S. Department of Energy, the National
Aeronautics and Space Administration, the Japanese Monbukagakusho, the
Max Planck Society, and the Higher Education Funding Council for
England. The SDSS Web Site is http://www.sdss.org/.

The SDSS is managed by the Astrophysical Research Consortium for the
Participating Institutions. The Participating Institutions are the
American Museum of Natural History, Astrophysical Institute Potsdam,
University of Basel, University of Cambridge, Case Western Reserve
University, University of Chicago, Drexel University, Fermilab, the
Institute for Advanced Study, the Japan Participation Group, Johns
Hopkins University, the Joint Institute for Nuclear Astrophysics, the
Kavli Institute for Particle Astrophysics and Cosmology, the Korean
Scientist Group, the Chinese Academy of Sciences (LAMOST), Los Alamos
National Laboratory, the Max-Planck-Institute for Astronomy (MPIA),
the Max-Planck-Institute for Astrophysics (MPA), New Mexico State
University, Ohio State University, University of Pittsburgh,
University of Portsmouth, Princeton University, the United States
Naval Observatory, and the University of Washington.

\end{document}